# Possibility of anapole state in dielectric nanohole array metasurfaces with different hole shapes

A. V. Panov(https://orcid.org/0000-0002-9624-4303)[a,*]

[a] *Institute of Automation and Control Processes, Far Eastern Branch of Russian Academy of Sciences, 5 Radio St., Vladivostok, 690041, Russia*

*\*e-mail: panov@iacp.dvo.ru*

**Abstract**—The optical anapole resonances in nanostructures display strong field confinement and substantially suppressed scattering. In this study using three-dimensional finite-difference time-domain simulations, it is shown that high refractive index dielectric nanohole array metasurfaces having different profiles of the holes can possess the anapole state. The multipole decomposition including the dipole electric toroidal moment for the lattice elements of the arrays is provided. The anapole state in the lattice elements is illustrated by time-averaged distributions of the energy. As a result, high-index freestanding metasurfaces with anapole state can be designed.

During the past years, significant attention by researchers has been given to the anapole states of high-index all-dielectric nanostructures. In the nanostructures, the anapole modes arise from destructive interference of the electric and toroidal dipole resonant modes of a specific charge-current distribution which display strong field enhancement along with reduced scattering [1]. Typically, the anapole states are observed in standalone dielectric nanoobjects (disks, spheres, cuboids etc.). Recently, high-index dielectric metasurfaces with circular nanopores were shown by the three-dimensional finite-difference time-domain (FDTD) simulations to possess the anapole state resulting in the effective optical Kerr nonlinearity increase by orders of magnitude [2]. This metasurface can be realized as a freestanding dielectric metasurface (membrane) that can be elaborated now for visible and near-infrared ranges [3,4]. Particularly, for near-infrared range this freestanding metasurface may be fabricated from silicon.

As shown in [2] by the multipole decomposition of scattering cross sections, for the specific geometric parameters of the circular nanohole array metasurface the maximum of the dipole electric toroidal moment is observed. This peak coincides with the minimum of the total scattering and can be illustrated by the double toroidal distribution of electric field energy. The transmission spectra of the circular nanohole arrays display high transmission in the vicinity to the anapole state. The optical nonlinearity is enhanced by two orders of magnitude than that of the unstructured material in proximity to the anapole state due to the energy confinement within the nanostructure.

In this study, the influence of the nanopore shape on the anapole state of the array lattice element is investigated. The base wavelength for the simulations is selected as λ=1034 nm that is typical for the Yb:YAG lasers. Figure 1 depicts schematics of the the nanohole square and hexagonal array metasurface. Here $b_4$ or $b_6$ are the sides of the nanohole, $a$=500 nm is the lattice constant, $h$ is the thickness of the metasurface which is equal to 200 nm in this work. The simulated linearly polarized Gaussian beam falls perpendicularly on the metasurface. The electric field of the incident beam is polarized along the $x$-axis.

After FDTD simulations, there were found geometric parameters for elements of the lattice with $a$=500 nm and $h$=200 nm for the square, hexagonal and octagonal nanopores. The lattice elements are delineated by the blue-lined square in Fig. 1. Figure 2 illustrates the time-averaged electric $|\mathbf{E}|^2$ and magnetic $|\mathbf{H}|^2$ energy distributions at transverse section of the lattice element at $h/2$ thickness. The electric energy distributions have shapes consisting of two loops giving rise to the anapole state. These energy distributions are like those obtained for the circular nanohole arrays in Ref. [2]. The lattice elements of the square and octagonal shapes of the nanoholes have approximately two times larger electric energy enhancement near the edges of the pore than that of the hexagonal shape. But it seems that this could not be realized in practice because of the imperfectness of the nanotechnology.

The existence of the anapole mode can be confirmed by the multipole analysis. A detailed multipole decomposition obtained after three-dimensional FDTD simulations of the scattered fields is presented in Fig. 3. The multipole decomposition of scattering cross sections for a lattice element reveals that the dipole electric toroidal moment $\mathbf{T}$ and its intensity $C^T$ has a maximum in the scattering cross section spectrum near the wavelength of interest 1034 nm at these sizes. Simultaneously, the total scattering cross section $C_{sca}^{tot}$ and the electric dipole cross section $C_{sca}^{p}$ have minima at $\lambda$=1034 nm that is the scattering from the lattice element is suppressed. The description of the decomposition moments and the procedure is given in [2]. All the spectra are very similar.

The transmission spectra of the Si nanohole arrays with $a$=500 nm and $h$=200 nm are depicted in Fig. 4. For wavelengths above the anapole state the transmission rapidly arises. This is a typical behavior for the anapole mode. The similar behavior of the transmission was observed for the lattices of circular nanoholes [2]. The dip in proximity to the anapole state corresponds to a spike in the scattering and the negative second order refractive index [2]. Since the transmission near the anapole state for Si nanohole array is below unity the optimal geometric parameters of the metasurface are yet to be determined before its implementation on the experiment. For example, GaP nanohole arrays show sharper increase in the transmission at the anapole state [2].

As can be seen from Figures 2–4, the shape of the nanopore has a minor effect on the existence of the anapole mode. Mostly, the geometric parameters for the anapole state are changed. The areas of the nanohole transverse sections $S$ for all of the shapes at the anapole state are close: for square $S$ = 0.085 μm$^2$, for hexagon $S$ = 0.102 μm$^2$, for octagon $S$ = 0.098 μm$^2$. Thus, it can be implemented in the manufactured metasurfaces which may be far from ideal geometric shapes due to the imperfectness of the technology.

In conclusion, the effect of the nanopore shape of the silicon nanohole lattice array on the existence of the electric dipole toroidal mode is studied. It is shown that all the investigated nanohole shapes (square, hexagonal and octagonal) allow the possibility of the anapole state with high electromagnetic energy confinement.

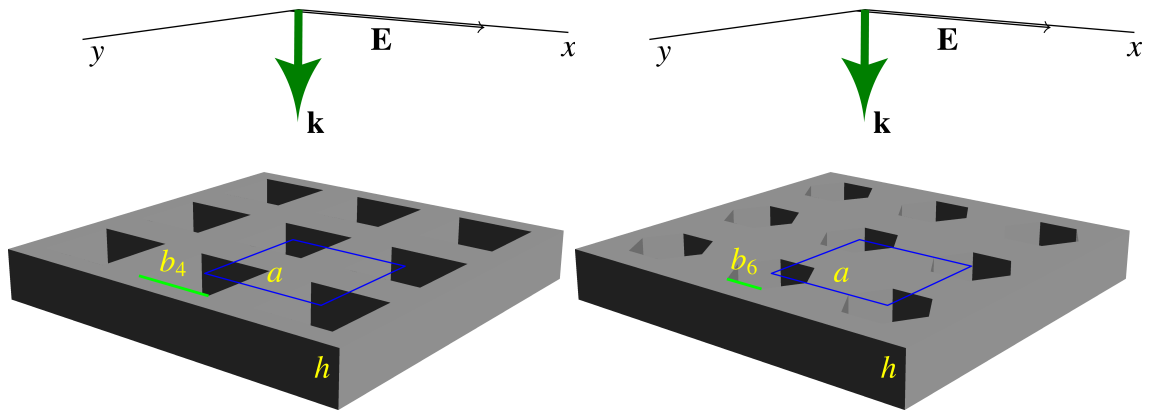

**Fig. 1.** Schematics of the simulated metasurfaces comprising a lattice arrays of square or hexagonal nanoholes in a high refractive index slab. The Gaussian beam is incident normally on the metasurface.

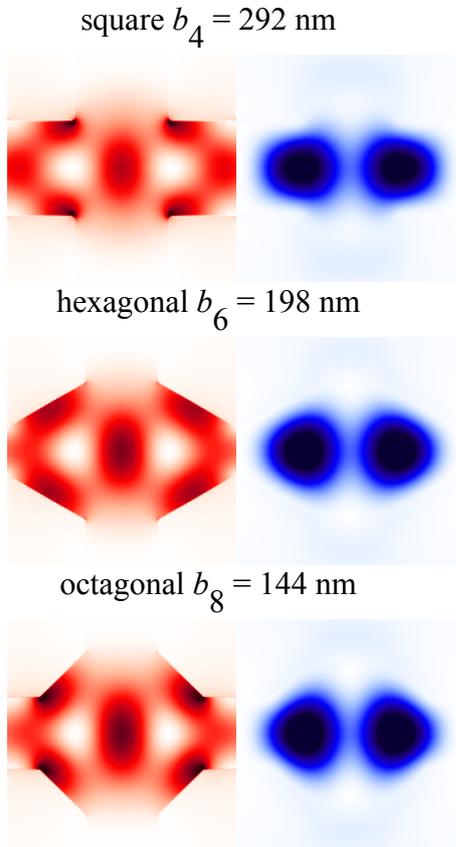

**Fig. 2.** Time-average distributions of electric $|\mathbf{E}|^2$ (left parts, red color) and magnetic $|\mathbf{H}|^2$ (right parts, blue color) energy densities in the standalone Si lattice elements at the anapole modes ($a$=500 nm, $h$=200 nm, $\lambda$=1034 nm) for the different types of the nanopores. The types of the nanohole and the polygon side sizes are displayed above. The distributions are calculated within the plane intersecting geometric center of the disk and parallel to its base. The incident light beam is polarized along the vertical direction.

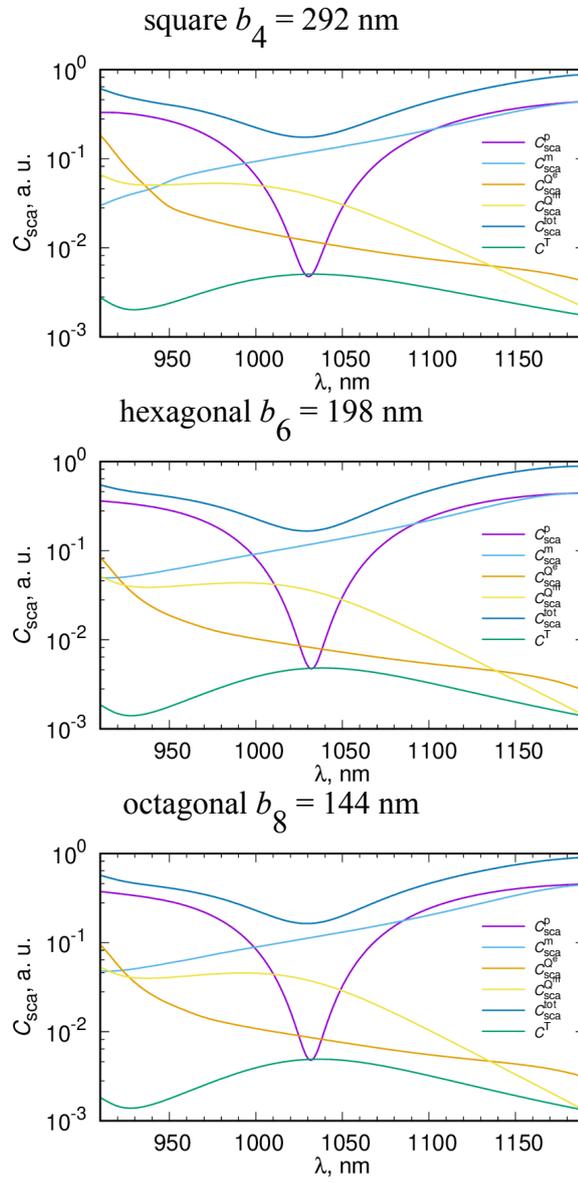

**Fig. 3.** Scattering cross section spectra for the multipole contributions (electric dipole $C_{sca}^{p}$, magnetic dipole $C_{sca}^{m}$, electric quadrupole $C_{sca}^{Q^e}$, magnetic quadrupole $C_{sca}^{Q^m}$), their sum $C_{sca}^{tot}$ and the intensity of the electric dipole toroidal moment $C^T$ for different Si lattice elements with $a$=500 nm, $h$=200 nm. The types of the nanohole and the nanohole transverse section side sizes are displayed above.

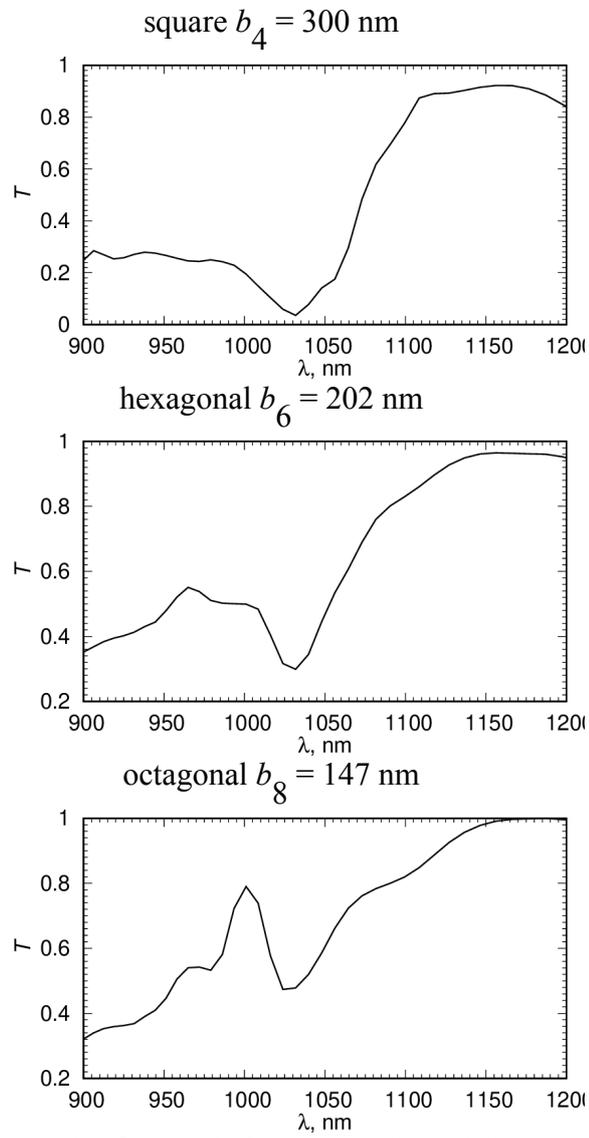

**Fig. 4.** Transmission spectra for the Si nanohole arrays of the different shape with $a$=500 nm and $h$=200 nm. The types of the nanohole and the side sizes are displayed above.